\documentclass[11pt,english]{article}
\usepackage{graphics, epsfig}

\title{Null Hypersurfaces in de Sitter and anti--de Sitter Cosmologies}

\author{P. A. Hogan\footnote{Email: peter.hogan@ucd.ie}, \\
        School of Physics\\
        University College Dublin\\
        Belfield, Dublin 4, Ireland}

\date{}

\begin{document}
\maketitle

\begin{abstract}
The study of gravitational waves in the presence of a cosmological constant has led to interesting forms of 
the de Sitter and anti--de Sitter line elements based on families of null hypersurfaces. The forms are interesting because they focus attention on the geometry of null hypersurfaces in space--times of constant curvature. 
Two examples are worked out in some detail. The first originated in the study of collisions of impulsive gravitational waves in which the post collision space--time is a solution of Einstein's field 
equations with a cosmological constant and the second originated in the generalisation of plane fronted gravitational waves with parallel rays to include a cosmological constant.
\end{abstract}

\setcounter{equation}{0}
\section{Introduction}
Some interesting forms of the de Sitter and anti--de Sitter line elements have emerged from the study of gravitational radiation in the presence of a cosmological constant. The forms are interesting because 
they focus attention on the geometry of null hypersurfaces in space--times of constant curvature. In this paper we borrow from two radiation studies of this type: the space--time of constant curvature which describes 
the gravitational field following a collision of two impulsive gravitational waves \cite{BH} and the space--time of constant curvature which is a spin--off from the Ozsv\'ath--Robinson--R\'ozga \cite{ORR} generalisation 
of the plane fronted gravitational waves with parallel rays to include a cosmological constant. In the first of these examples, discussed in section 2 below, the null hypersurfaces that emerge have 
non--vanishing shear and expansion and are thus quite complicated objects. In the second example the null hypersurfaces are null hyper{\it planes} generated by shear--free and expansion--free null geodesics. 
These geometrical objects must not be confused with the simpler null hypersurfaces generated by the integral curves of a covariantly constant null vector field. It is these latter objects which are more closely analogous 
to the histories of homogeneous plane waves in Maxwell's electromagnetic theory. To 
discuss the null hyperplanes in sections 3 and 4 below we exploit the conformally flat property of space--times of constant curvature, in particular making use of the fact that being null, geodesic and shear--free are conformally invariant 
concepts. We also exploit the fact that being expansion--free is \emph{not} a conformally invariant property. The line elements of de Sitter and anti--de Sitter space--times which can be found in \cite{ORR} 
have the unusual feature of depending upon three real--valued functions of a null coordinate (in the form of a real--valued function and the real and imaginary parts of a complex--valued function). The 
geometrical construction described in detail in sections 3 and 4 provides a clear explanation of the origin of these functions. Finally in section 5, using the formalism developed in the two previous sections, we establish a property of the null hyperplanes, which was originally discovered by Tran and Robinson \cite{Tran}, \cite{Rob} (see also \cite{BH1}) using the representation of a space--time of constant curvature as a quadric in a five dimensional flat space--time, which representation we make no use of here.

A fascinating description of the history of de Sitter and anti--de Sitter space--times has been given by Schucking and Wang \cite{SW}. The representation of these space--times as quadrics in a five dimensional flat 
space--time can be found in \cite{Sch}, \cite{Syn} and \cite{HE} for example. When different coordinates are utilised various properties of the manifolds can easily be exhibited, for example, the existence of three dimensional 
submanifolds of constant positive curvature or of zero curvature in the case of de Sitter space--time (see \cite{HE}, p.125). In the present paper the focus is on null hypersurfaces in these space--times and how coordinates based 
on families of such objects help to illustrate properties of them. The process of finding coordinate representations of de Sitter and anti--de Sitter space--times to illustrate aspects of these manifolds continues to the present day 
with the most recent known example found in \cite{PH}. 

\setcounter{equation}{0}
\section{Null Hypersurfaces with Shear}
The head--on collision of two plane and homogeneous impulsive gravitational waves can lead to a post collision de Sitter or anti--de Sitter \cite{BH} space--time with a novel form of line element expressed in 
a coordinate system based on two families of intersecting null hypersurfaces $u={\rm constant}$ and $v={\rm constant}$ (say). This line element is 
\begin{equation}\label{1'}
ds^2=\frac{(1-k\,l\,u\,v+k\,u+l\,v)^2dx^2+(1-k\,l\,u\,v-k\,u-l\,v)^2dy^2-2\,du\,dv}{(1+k\,l\,u\,v)^2}\ ,\end{equation}
where $k, l$ are real constants and the cosmological constant $\Lambda$ is given by
\begin{equation}\label{2'}
\Lambda=-6\,k\,l\ .\end{equation}The constants $k, l$ appear because the amplitudes of the incoming waves were proportional to $k$ and $l$ respectively. The post collision field equations are chosen to involve a cosmological constant. The Riemann curvature tensor components, calculated with the metric tensor given via the line element (\ref{1'}), take the form
\begin{equation}\label{3'}
R_{ijkl}=2\,k\,l\,(g_{il}\,g_{jk}-g_{ik}\,g_{jl})\ .\end{equation}We note that with the inverse of the metric tensor having components $g^{ij}$ (and hence with $g^{ij}\,g_{jk}=\delta^i_k$), and a comma denoting 
partial differentiation with respect to the coordinates $x^i$,  we have $g^{ij}u_{,i}\,u_{,j}=0=g^{ij}v_{,i}\,v_{,j}$ and $g^{ij}u_{,i}\,v_{,j}=-(1+k\,l\,u\,v)^2\neq 0$ confirming that $u={\rm constant}$ and $v={\rm constant}$ 
are intersecting families of null hypersurfaces. These are not the simplest null hypersurfaces that one can find in de Sitter space--time (when $\Lambda>0\ \Leftrightarrow k\,l<0$) or in anti--de Sitter space--time (when $\Lambda<0\ \Leftrightarrow k\,l>0$). Their 
null geodesic generators have non--vanishing shear and expansion. To see this define a null tetrad via the 1--forms
\begin{eqnarray}
m_i\,dx^i&=&\frac{\{(1-k\,l\,u\,v)(dx+idy)+(k\,u+l\,v)(dx-idy)\}}{(1+k\,l\,u\,v)}\ ,\label{4'}\\
\bar m_i\,dx^i&=&\frac{\{(1-k\,l\,u\,v)(dx-idy)+(k\,u+l\,v)(dx+idy)\}}{(1+k\,l\,u\,v)}\ ,\label{5'}\\
k_i\,dx^i&=&(1+k\,l\,u\,v)^{-1}du\ ,\label{6'}\\
l_i\,dx^i&=&(1+k\,l\,u\,v)^{-1}dv\ ,\label{7'}\end{eqnarray}
with $i=\sqrt{-1}$ and the bar denoting complex conjugation. The integral curves of $k_i$ generate the null hypersurfaces $u={\rm constant}$ and the integral curves 
of $l_i$ generate the null hypersurfaces $v={\rm constant}$. The shear and expansion of $k_i$ are given respectively by
\begin{equation}\label{8'}
\sigma_k=k_{i;j}\,m^i\,m^j=-\frac{2\,l(1+k\,l\,u\,v)(1+k^2u^2)}{(1-k\,l\,u\,v)^2-(k\,u+l\,v)^2}\ ,\end{equation}
and
\begin{equation}\label{9'}
\rho_k=k_{i;j}\,m^i\,\bar m^j=\frac{2\,l\,\{k\,u(3-k^2u^2)+l\,v(1-3\,k^2u^2)\}}{(1-k\,l\,u\,v)^2-(k\,u+l\,v)^2}\ ,\end{equation}
with the semicolon denoting covariant differentiation with respect to the Riemannian connection calculated with the metric tensor $g_{ij}$. Similarly the 
shear and expansion of $l_i$ are given by
\begin{equation}\label{10'}
\sigma_l=l_{i;j}\,m^i\,m^j=-\frac{2\,k(1+k\,l\,u\,v)(1+l^2v^2)}{(1-k\,l\,u\,v)^2-(k\,u+l\,v)^2}\ .\end{equation}
and
\begin{equation}\label{11'}
\rho_l=l_{i;j}\,m^i\,\bar m^j=\frac{2\,k\,\{l\,v(3-l^2v^2)+k\,u(1-3\,l^2v^2)\}}{(1-k\,l\,u\,v)^2-(k\,u+l\,v)^2}\ .\end{equation}
The essential feature of the optical scalars (\ref{8'}) and (\ref{9'}) associated with the null hypersurfaces  $u={\rm constant}$ and the optical scalars (\ref{10'}) and 
(\ref{11'}) associated with the null hypersurfaces $v={\rm constant}$ is that these scalars are in general non--vanishing. The de Sitter and anti--de Sitter space--times admit 
null hypersurfaces having the property that their associated optical scalars vanish (and so the hypersurfaces are shear--free and expansion--free). These latter null hypersurfaces 
lend themselves to more detailed study and this is carried out in sections 3, 4, and 5 below. However notwithstanding the complexity of the intersecting null hypersurfaces $u={\rm constant}$ and $v={\rm constant}$, we can still make interesting associations between the 
new form (\ref{1'}) of the de Sitter or anti--de Sitter line element and more familiar forms. We turn to this now.

Making the coordinate transformation
\begin{equation}\label{12'}
k\,u=\tan\left (U-\frac{\pi}{8}\right )\ ,\ \ l\,v=\tan\left (V-\frac{\pi}{8}\right )\ ,\end{equation}
results in (\ref{1'}) taking the form
\begin{equation}\label{13'}
ds^2=\sec^2(U-V)\left\{2\,\sin^2(U+V)\,dx^2+2\,\cos^2(U+V)\,dy^2-\frac{2}{k\,l}dU\,dV\right\}\ .\end{equation}
We must now consider separately the special cases of de Sitter space--time and anti--de Sitter space--time. 

We begin with the de Sitter case and write for convenience 
\begin{equation}\label{14'}
\epsilon_0^2=-2\,k\,l=\frac{\Lambda}{3}>0\ .\end{equation}Putting first
\begin{equation}\label{15'}
x=X/\epsilon_0\sqrt{2}\ ,\ \ y=Y/\epsilon_0\sqrt{2}\ ,\ U+V=Z\ \ {\rm and}\ \ U-V=\eta\ ,\end{equation}
and then writing 
\begin{equation}\label{16'}
\sin\eta=\tanh\epsilon_0T\ ,\end{equation}transforms (\ref{13'}) into
\begin{equation}\label{17'}
ds^2=\frac{\cosh^2\epsilon_0T}{\epsilon_0^2}\left (\sin^2Z\,dX^2+\cos^2Z\,dY^2+dZ^2\right )-dT^2\ .\end{equation}
This is a standard form of the de Sitter line element \cite{HE} in which the spatial sections $T={\rm constant}$ are 3--spheres as can be seen by restricting the line element 
of four dimensional Euclidean space, in rectangular Cartesian coordinates $(z^1, z^2, z^3, z^4)$,
\begin{equation}\label{18'}
ds_0^2=(dz^1)^2+(dz^2)^2+(dz^3)^2+(dz^4)^2\ ,\end{equation}
to points on the 3--sphere with equation
\begin{equation}\label{19'}
(z^1)^2+(z^2)^2+(z^3)^2+(z^4)^2=1\ .\end{equation}
With the parametrisation
\begin{equation}\label{20'}
z^1+iz^2=e^{iX}\sin Z\ ,\ z^3+iz^4=e^{iY}\cos Z\ ,\end{equation}
(\ref{19'}) is satisfied and (\ref{18'}) reduces to 
\begin{equation}\label{21'}
ds_0^2=\sin^2Z\,dX^2+\cos^2Z\,dY^2+dZ^2\ .\end{equation}

For the anti--de Sitter case we write
\begin{equation}\label{22'}
\epsilon_0^2=2\,k\,l=-\frac{\Lambda}{3}>0\ ,\end{equation}
and transform (\ref{13'}) with
\begin{equation}\label{23'}
x=X/\epsilon_0\sqrt{2}\ ,\ \ y=Y/\epsilon_0\sqrt{2}\ ,\ U+V=T\ \ {\rm and}\ \ U-V=\eta\ ,\end{equation}
followed by
\begin{equation}\label{24'}
\sin\eta=\tanh\epsilon_0Z\ ,\end{equation}
to arrive at
\begin{equation}\label{25'}
ds^2=\frac{\cosh^2\epsilon_0Z}{\epsilon_0^2}\left\{\sin^2T\,dX^2+\cos^2T\,dY^2-dT^2\right\}+dZ^2\ .\end{equation}
For this rather unusual form of the line element of anti--de Sitter space--time the spatial sections $T={\rm constant}$ are not homogeneous. However the 
sections $Z={\rm constant}$ do have constant (negative) curvature. They can be embedded in a four dimensional flat pseudo--Riemannian manifold 
of points $(z^1, z^2, z^3, z^4)$ having line element
\begin{equation}\label{26'}
ds_0^2=(dz^1)^2-(dz^2)^2+(dz^3)^2-(dz^4)^2\ .\end{equation}If the points $(z^1, z^2, z^3, z^4)$ are restricted to lie on the 3--surface
\begin{equation}\label{27'}
(z^1)^2-(z^2)^2+(z^3)^2-(z^4)^2=-1\ ,\end{equation}
given parametrically by
\begin{equation}\label{28'}
z^1=\cos T\,\sinh Y\ ,\ z^2=\sin T\,\cosh X\ ,\end{equation}
and
\begin{equation}\label{29'}
z^3=\sin T\,\sinh X\ ,\ z^4=\cos T\,\cosh Y\ ,\end{equation}
then (\ref{26'}) becomes
\begin{equation}\label{30'}
ds_0^2=\sin^2T\,dX^2+\cos^2T\,dY^2-dT^2\ .\end{equation}

\setcounter{equation}{0}
\section{Null Hyperplanes}\indent 
A null hyperplane is a null hypersurface generated by null geodesics which are shear--free and expansion--free (and, of course, twist--free). The properties of being \emph{null}, \emph{geodesic} and 
\emph{shear--free} are conformally invariant properties and thus help to construct null hyperplanes in space--times of constant curvature since such space--times are conformally flat. In addition all the shear--free, null 
hypersurfaces in flat space--time are known. They are either null hyperplanes or null cones or portions thereof \cite{P}. Let $X^i=(X, Y, Z, T)$ be rectangular Cartesian coordinates and time in Minkowskian space--time 
with line element
\begin{equation}\label{1}
ds^2=dX^2+dY^2+dZ^2-dT^2=\eta_{ij}dX^i\,dX^j\ ,\end{equation}
with $\eta_{ij}={\rm diag}(1, 1, 1, -1)$ the components of the Minkowskian metric tensor in coordinates $X^i$. Latin indices take values 1, 2, 3, 4 and indices are raised and lowered using $\eta^{ij}$ and $\eta_{ij}$ 
respectively with $\eta^{ij}$ defined by $\eta^{ij}\eta_{jk}=\delta^i_k$. Shear--free null hypersurfaces in Minkowskian space--time are given by $u(X, Y, Z, T)={\rm constant}$, with $u(X, Y, Z, T)$ defined implicitly 
by the equation of a null hyperplane:
\begin{equation}\label{2}
\eta_{ij}a^i(u)X^j+b(u)=0\ \ \ {\rm with}\ \ \eta_{ij}a^ia^j=0\ ,\end{equation}
or the equation of a null cone with vertex on the arbitrary line $X^i=w^i(u)$:
\begin{equation}\label{3}
\eta_{ij}(X^i-w^i(u))(X^j-w^j(u))=0\ .\end{equation}

We first verify that $u={\rm constant}$ given by (\ref{2}) are null hyper{\rm planes}. Differentiating (\ref{2}) with respect to $X^k$ yields
\begin{equation}\label{4}
u_{,k}=-\varphi^{-1}a_k\ \ {\rm with}\ \ \ \varphi=\dot b+\dot a_i\,X^i\ ,\end{equation}
and the partial derivative is denoted by a comma. The dot denotes differentiation with respect to $u$. Since $a^i$ is a null vector field this confirms that $u={\rm constant}$ are null hypersurfaces. Differentiating (\ref{4}) with respect 
to $X^l$ results in
\begin{equation}\label{5}
u_{,kl}=-\varphi^{-1}(\dot a_k\,u_{,l}+\dot a_l\,u_{,k})-\varphi^{-1}\dot\varphi\,u_{,k}\,u_{,l}\ .\end{equation}From the algebraic form of the right hand side of this equation \cite{RobTra} it follows that 
the covariant null vector field $u_{,k}$ is geodesic and shear--free (it is obviously twist--free). From (\ref{5}) we deduce that
\begin{equation}\label{6}
\eta^{kl}u_{,kl}=-2\,\varphi^{-1}\eta^{kl}\dot a_k\,u_{,l}=2\,\varphi^{-2}\eta^{kl}\dot a_k\,a_l=0\ ,\end{equation}
since $a^i$ is null. Hence, in addition to being {\it null}, {\it geodesic}, {\it shear--free} and {\it twist--free}, $u_{,k}$ is also {\it expansion--free}. Thus the null hypersurfaces $u={\rm constant}$ are 
null hyper{\it planes}.

The space--times of constant (non--zero) curvature are de Sitter space--time (positive curvature) or anti--de Sitter space--time (negative curvature) depending upon the sign of the cosmological constant $\Lambda$. These 
space--times are conformally flat and the line element can be written in the conformally flat form:
\begin{equation}\label{7}
ds^2=\lambda^2\eta_{ij}dX^i\,dX^j\ ,\end{equation}
with
\begin{equation}\label{8}
\lambda=\left (1+\frac{\Lambda}{12}\eta_{ij}X^i\,X^j\right )^{-1}\ .\end{equation}From the conformal invariance of the null, geodesic and shear--free properties we know that $u={\rm constant}$ given by (\ref{2}) 
are null, geodesic and shear--free in the space--time with line element (\ref{7}). We now look for the condition that $u={\rm constant}$ are expansion--free in the space--time with line element (\ref{7}). For this 
we must calculate $u_{,k;l}$ with the semicolon indicating covariant differentiation with respect to the Riemannian connection associated with the metric tensor $g_{ij}=\lambda^2\eta_{ij}$. The components of 
this Riemannian connection are
\begin{equation}\label{9}
\Gamma^i_{jk}=\lambda^{-1}(\lambda_{,j}\,\delta^i_k+\lambda_{,k}\,\delta^i_j-\eta^{ip}\lambda_{,p}\,\eta_{jk})\ .\end{equation} Thus we find, using (\ref{5}), that
\begin{eqnarray}
u_{,k;l}&=&-(\varphi^{-1}\dot a_k+\lambda^{-1}\lambda_{,k})\,u_{,l}-(\varphi^{-1}\dot a_l+\lambda^{-1}\lambda_{,l})\,u_{,k}\nonumber\\
&&-\varphi^{-1}\dot\varphi\,u_{,k}\,u_{,l}+\lambda^{-1}\eta^{pq}\lambda_{,p}\,u_{,q}\,\eta_{kl}\ .\label{10}\end{eqnarray} This again has the correct algebraic form \cite{RobTra} for $u_{,k}$ to be 
geodesic and shear--free in the space--time with line element (\ref{7}). From (\ref{10}) we see that
\begin{equation}\label{11}
\eta^{kl}u_{,k;l}=2\,\lambda^{-1}\eta^{pq}\lambda_{,p}\,u_{,q}=-2\,\lambda^{-1}\varphi^{-1}\eta^{pq}\lambda_{,p}\,a_q\ .\end{equation}But $\lambda_{,p}=-\lambda^2\Lambda\,\eta_{pq}X^q/6$ and so
\begin{equation}\label{12}
\eta^{pq}\lambda_{,p}\,a_q=-\frac{\Lambda}{6}\lambda^2a_p\,X^p=\frac{\Lambda}{6}\lambda^2b\ ,\end{equation}using (\ref{2}). Hence the null hyperplanes (\ref{2}) in Minkowskian space--time are 
null hyper{\it planes} in the space--time with line element (\ref{7}) provided $b=0$ (which, by (\ref{11}) and (\ref{12}), is necessary in order to have $u_{,k}$ expansion--free in the space--time with line element (\ref{7})). Thus  
the null hyperplanes $u={\rm constant}$ in Minkowskian space--time given by
\begin{equation}\label{13}
a_i(u)\,X^i=0\ ,\end{equation}
correspond to null hyperplanes in the space--time of constant curvature with line element (\ref{7}). The null hyperplanes (\ref{13}) pass through the origin $X^i=0$ and are tangent to the null cone with vertex $X^i=0$, and therefore intersect each other. Only the \emph{direction} of the null vector field $a^i$ is significant in (\ref{13}) and this is determined by two real--valued functions of $u$ or equivalently by one complex--valued function $l(u)$ with 
complex conjugate $\bar l(u)$. Thus we can write
\begin{equation}\label{14}
a^1+ia^2=2\,\sqrt{2}\,l\ ,\ \ a^3+a^4=4\,l\bar l\ ,\ \ a^3-a^4=-2\ .\end{equation}Now (\ref{13}) reads:
\begin{equation}\label{15}
Z+T=\sqrt{2}\,\bar l(X+iY)+\sqrt{2}\,l(X-iY)+2\,l\bar l(Z-T)\ .\end{equation}Using this we find that
\begin{equation}\label{16}
\eta_{ij}X^i\,X^j=\left |X+iY+\sqrt{2}\,l\,(Z-T)\right |^2\ ,\end{equation}which suggests that we introduce a complex coordinate $\zeta$ (with complex conjugate denoted by a bar) via
\begin{equation}\label{17}
\zeta=\frac{1}{\sqrt{2}}(X+iY)+l\,(Z-T)\ .\end{equation}Now instead of using the coordinates $X, Y, Z, T$ we may use coordinates $\zeta, \bar\zeta, u, Z-T$ satisfying
\begin{eqnarray}
X+iY&=&\sqrt{2}\,\zeta-\sqrt{2}\,l\,(Z-T)\ ,\label{18}\\
Z+T&=&2\,(\bar l\,\zeta+l\,\bar\zeta)-2\,l\,\bar l(Z-T)\ ,\label{19}\end{eqnarray}while the conformal factor $\lambda$ is given using (\ref{8}) and (\ref{16}) by
\begin{equation}\label{19}
\lambda^{-1}=1+\frac{\Lambda}{12}\eta_{ij}X^i\,X^j=1+\frac{\Lambda}{6}\zeta\bar\zeta=p\ {\rm (say)}\ .\end{equation}Now the line element (\ref{7}) reads
\begin{equation}\label{20}
ds^2=2\,p^{-2}d\zeta\,d\bar\zeta+2\,p^{-2}d\Sigma\,du ,\end{equation}where $d\Sigma$ (which is not necessarily an exact differential) is given by
\begin{equation}\label{21}
d\Sigma=-(Z-T)(\beta\,d\bar\zeta+\bar\beta\,d\zeta)+(\beta\,\bar\zeta+\bar\beta\,\zeta)(dZ-dT)+\beta\bar\beta(Z-T)^2du\ ,\end{equation}
where $\beta(u)=dl(u)/du$. If we now define
\begin{equation}\label{22}
q=\beta\,\bar\zeta+\bar\beta\,\zeta\ ,\end{equation}and in place of $Z-T$ use a coordinate $r$ defined by
\begin{equation}\label{23}
Z-T=q\,r\ ,\end{equation}
then the line element (\ref{20}) takes the Ozsv\'ath-Robinson-R\'ozga \cite{ORR} form
\begin{equation}\label{24}
ds^2=2\,p^{-2}d\zeta\,d\bar\zeta+2\,p^{-2}q^2du\{dr+(q^{-1}\dot q\,r+\beta\bar\beta\,r^2)du\}\ ,\end{equation}
where the dot, as always, denotes differentiation with respect to $u$. This is a special case of this form of line element. The more general case emerges in the next section.

\setcounter{equation}{0}
\section{From Null Cones to Null Hyperplanes}\indent
Writing
\begin{equation}\label{25}
\xi^i=X^i-w^i(u)\ ,\end{equation}
we have from (\ref{3}) that
\begin{equation}\label{26}
\eta_{ij}\xi^i\,\xi^j=0\ .\end{equation}Differentiating this with respect to $X^k$ results in 
\begin{equation}\label{27}
u_{,k}=\frac{\xi_k}{R}=k_k\ \ ({\rm say})\ ,\end{equation}
with
\begin{equation}\label{28}
R=\eta_{ij}\dot w^i\,\xi^j\ \ {\rm and}\ {\rm thus}\ \eta_{ij}\dot w^i\,k^j=+1\ .\end{equation}As always a dot indicates differentiation with respect to $u$. It is clear from (\ref{26}) and (\ref{27}) 
that the hypersurfaces $u={\rm constant}$ are null. Straightforward calculations yield
\begin{equation}\label{29}
\xi^i{}_{,j}=\delta^i_j-\dot w^i\,k_j\ ,\end{equation}
and
\begin{equation}\label{30}
R_{,i}=\dot w_i+A\,k_i\ ,\end{equation}
with $\dot w_i=\eta_{ij}\dot w^j$ and 
\begin{equation}\label{31}
A=-\dot w_i\,\dot w^i+R\,\ddot w_i\,k^i\ .\end{equation}
Using (\ref{27})--(\ref{31}) we arrive at
\begin{equation}\label{32}
k_{i,j}=\frac{1}{R}(\eta_{ij}-\dot w_i\,k_j-\dot w_j\,k_i-A\,k_i\,k_j)=k_{j,i}\ ,\end{equation}
which displays the algebraic structure \cite{RobTra} guaranteeing that $k_i$ is geodesic and shear--free with expansion $\vartheta=k^i{}_{,i}/2=1/R\neq 0$. With a semicolon, as before, indicating 
covariant differentiation with respect to the Riemannian connection (\ref{9}) associated with the metric tensor given via the line element (\ref{7}) we find that
\begin{eqnarray}
k_{i;j}&=&(R^{-1}+\lambda^{-1}\eta^{kl}\lambda_{,k}\,k_l)\eta_{ij}-(R^{-1}\dot w_i+\lambda^{-1}\lambda_{,i})k_j\nonumber\\
&&-(R^{-1}\dot w_j+\lambda^{-1}\lambda_{,j})k_i-A\,R^{-1}k_i\,k_j\ .\label{33}\end{eqnarray}The algebraic form of this \cite{RobTra} ensures that $k_i$ is geodesic and shear--free in the space--time with line element (\ref{7}). The expansion vanishes 
if $k^i{}_{;i}$ vanishes. It follows from (\ref{33}) that this condition reduces to
\begin{equation}\label{34}
R^{-1}+\lambda^{-1}\lambda_{,i}\,k^i=0\ .\end{equation}With $\lambda$ given by (\ref{8}) this becomes
\begin{equation}\label{35}
\eta_{ij}w^i\,w^j=-\frac{12}{\Lambda}\ .\end{equation}

Since $w^i(u)$ has three independent components, a convenient parametrisation in terms of the real--valued function $m(u)$ and the 
complex--valued function $l(u)$ (with complex conjugate denoted $\bar l(u)$) is given by
\begin{equation}\label{36}
w^1+iw^2=\frac{6\,\sqrt{2}\,l}{\Lambda\,m}\ ,\ w^3+w^4=\frac{6}{\Lambda\,m}\left (\frac{1}{3}\Lambda\,m^2+2\,l\,\bar l\right )\ ,\ w^3-w^4=-\frac{6}{\Lambda\,m}\ .\end{equation}Here we assume that 
$m\neq 0$ but $m$ is small then (\ref{26}) approximates (\ref{13}) with $a^i$ given by (\ref{14}) and so we can expect that the results of this section will include those of the previous 
section in the limit of small $m(u)$. Writing out (\ref{3}) with $w^i(u)$ given by (\ref{36}) results in 
\begin{eqnarray}
Z+T&=&\sqrt{2}\,\bar l\,(X+iY)+\sqrt{2}\,l\,(X-iY)+2\,\l\,\bar l\,(Z-T)\nonumber\\
&&+2\,m\left (1+\frac{\Lambda}{6}\,m\,(Z-T)\right )-\frac{\Lambda}{6}\,m\,\eta_{ij}X^i\,X^j\ ,\label{37}\end{eqnarray}
which specialises to (\ref{15}) when $m=0$. Using this we can write
\begin{eqnarray}
\left (1+\frac{\Lambda}{6}\,m\,(Z-T)\right )\eta_{ij}\,X^iX^j&=&\left |X+iY+\sqrt{2}\,l\,(Z-T)\right |^2\nonumber\\
&&+2\,m\,(Z-T)\left (1+\frac{\Lambda}{6}\,m\,(Z-T)\right )\ .\nonumber\\\label{38}\end{eqnarray}This specialises to (\ref{16}) when $m=0$. From this we see that
\begin{eqnarray}
\left (1+\frac{\Lambda}{6}\,m\,(Z-T)\right )\lambda^{-1}&=&\left (1+\frac{\Lambda}{6}\,m\,(Z-T)\right )^2\nonumber\\
&&+\frac{\Lambda}{12}\left |X+iY+\sqrt{2}\,l\,(Z-T)\right |^2\ ,\label{39}\end{eqnarray}
with $\lambda$ given by (\ref{8}). In similar fashion to (\ref{17}) this suggests that we should define the new complex coordinate
\begin{equation}\label{40}
\zeta=\left (1+\frac{\Lambda}{6}\,m\,(Z-T)\right )^{-1}\left\{\frac{1}{\sqrt{2}}(X+iY)+l\,(Z-T)\right\}\ ,\end{equation}
so that
\begin{equation}\label{41}
\lambda^{-1}=\left (1+\frac{\Lambda}{6}\,m\,(Z-T)\right )\,p\ ,\end{equation}
with
\begin{equation}\label{42}
p=1+\frac{\Lambda}{6}\zeta\bar\zeta\ .\end{equation}Now instead of using the coordinates $X, Y, Z, T$ we shall use the coordinates $\zeta, \bar\zeta, u, Z-T$ with
\begin{eqnarray}
X+iY&=&\sqrt{2}\,\left (1+\frac{\Lambda}{6}\,m\,(Z-T)\right )\,\zeta-\sqrt{2}\,l\,(Z-T)\ ,\label{43}\\
Z+T&=&2\,\left\{\bar l\,\zeta+l\,\bar\zeta+m\,\left (1-\frac{\Lambda}{6}\zeta\bar\zeta\right )\right\}\left (1+\frac{\Lambda}{6}\,m\,(Z-T)\right )\nonumber\\
&&-\left (\frac{\Lambda}{3}\,m^2+2\,l\,\bar l\right )(Z-T)\ .\label{44}\end{eqnarray}Now the line element (\ref{7}) reads
\begin{equation}\label{45}
ds^2=2\,p^{-2}d\zeta\,d\bar\zeta+2\,p^{-2}d\Sigma\,du\ ,\end{equation}
with
\begin{eqnarray}
d\Sigma&=&\left (1+\frac{\Lambda}{6}\,m\,(Z-T)\right )^{-2}\Biggl\{\left (\frac{1}{2}\kappa-\frac{\Lambda}{6}\,\alpha\,q\right )(Z-T)^2du\nonumber\\
&&+q\,(dZ-dT)-(Z-T)\left (1+\frac{\Lambda}{6}\,m\,(Z-T)\right )\,(dq-\dot q\,du)\Biggr\}\ ,\nonumber\\\label{46}\end{eqnarray}
where $\alpha(u)=dm(u)/du, \beta(u)=dl(u)/du$ and
\begin{eqnarray}
\kappa&=&\frac{\Lambda}{3}\,\alpha^2+2\,\beta\,\bar\beta\ ,\label{47}\\
q(\zeta, \bar\zeta, u)&=&\beta\,\bar\zeta+\bar\beta\,\zeta+\alpha\,\left (1-\frac{\Lambda}{6}\zeta\,\bar\zeta\right )\ ,\label{48}\end{eqnarray}
and $\dot q=\partial q/\partial u$. If we now introduce the coordinate $r$ via the equation
\begin{equation}\label{49}
Z-T=\left (1+\frac{\Lambda}{6}\,m\,(Z-T)\right )\,q\,r\ ,\end{equation}the line element (\ref{45}) takes the general Ozsv\'ath--Robinson--R\'ozga \cite{ORR} form
\begin{equation}\label{50}
ds^2=2\,p^{-2}d\zeta\,d\bar\zeta+2\,p^{-2}q^2du\{dr+(q^{-1}\dot q\,r+\frac{1}{2}\kappa\,r^2)\,du\}\ ,\end{equation}with $p, q$ given by (\ref{42}) and (\ref{48}) respectively. 
When $m=0 (\Rightarrow \alpha=0)$ this reduces to (\ref{24}). The construction given here, and in the previous section, illustrates an origin for the arbitrary functions $\alpha(u), \beta(u)$ appearing in (\ref{50}) 
via (\ref{47}) and (\ref{48}).

\setcounter{equation}{0}
\section{Intersecting Null Hyperplanes}\indent
The equations of the null hyperplanes $u(X, Y, Z, T)={\rm constant}$, are given implicitly by (\ref{2}) with $b=0$ and by (\ref{3}) with (\ref{25}) holding. These are easy to visualise in Minkowskian space--time and 
so it is clear that the null hyperplanes given by (\ref{2}) with $b=0$ intersect. The null cones (\ref{3}) intersect if the world line $X^i=w^i(u)$ is space--like or time--like and they also intersect if this world line is, in general, null except 
when this null world line is a common generator of the null cones. For the latter to happen the world line $X^i=w^i(u)$ must be a null geodesic. With $w^i(u)$ given by (\ref{36}) we find that
\begin{equation}\label{51}
\eta_{ij}\dot w^i\,\dot w^j=\left (\frac{6}{\Lambda\,m}\right )^2\kappa\ ,\end{equation}with $\kappa$ given by (\ref{47}). Hence the character of the world line $X^i=w^i(u)$ depends upon the sign of $\kappa$. For $X^i=w^i(u)$ to be 
a null geodesic we must have $\kappa=0$ and
\begin{equation}\label{52}
\ddot w^i=C(u)\,\dot w^i\ ,\end{equation}for some real--valued function $C(u)$. Substituting (\ref{36}) into
\begin{equation}\label{53}
\ddot w^3-\ddot w^4=C\,(\dot w^3-\dot w^4)\ ,\end{equation}results in 
\begin{equation}\label{54}
C=\frac{1}{\alpha}\frac{d\alpha}{du}-\frac{2\,\alpha}{m}\ .\end{equation}Now (\ref{36}) in 
\begin{equation}\label{55}
\ddot w^1+i\ddot w^2=C(\dot w^1+i\dot w^2)\ ,\end{equation}along with (\ref{54}), produces
\begin{equation}\label{56}
\frac{1}{\beta}\frac{d\beta}{du}=\frac{1}{\alpha}\frac{d\alpha}{du}\ \ \Rightarrow\ \ \frac{1}{\beta}\frac{d\beta}{du}=\frac{1}{\bar\beta}\frac{d\bar\beta}{du}\ .\end{equation}As a consequence of (\ref{56}) the 
equation
\begin{equation}\label{57}
\ddot w^3+\ddot w^4=C\,(\dot w^3+\dot w^4)\ ,\end{equation}is automatically satisfied. Finally we note that 
\begin{equation}\label{58}
\kappa=0\ \ {\rm and}\ \ \frac{1}{\beta}\frac{d\beta}{du}=\frac{1}{\bar\beta}\frac{d\bar\beta}{du}\ \ \ \Rightarrow\ \ \ \frac{1}{\beta}\frac{d\beta}{du}=\frac{1}{\alpha}\frac{d\alpha}{du}\ ,\end{equation}and
\begin{equation}\label{59}
\frac{1}{\beta}\frac{d\beta}{du}=\frac{1}{\bar\beta}\frac{d\bar\beta}{du}\ \ \Rightarrow\ \ {\rm Re}\,\beta=0\ \ {\rm or}\ \ {\rm Im}\,\beta=0\ \ {\rm or}\ \ {\rm Re}\,\beta=c_0\,{\rm Im}\,\beta\ ,\end{equation}
for some real number $c_0$. We can summarise the results here in the theorem of Tran and Robinson \cite{Tran}, \cite{Rob}: (1) $\Lambda>0$\ \ $\Rightarrow\ \ \kappa>0\ \ \Rightarrow$\ \ intersecting null hyperplanes; 
(2) $\Lambda<0$\ \ $\Rightarrow\ \ \kappa>0$ or $\kappa<0$ or $\kappa=0$ with (i) $\kappa>0\ \ \Rightarrow$ intersecting null hyperplanes, (ii) $\kappa<0\ \ \Rightarrow$ intersecting null hyperplanes, (iii) $\kappa=0\ \ \Rightarrow$ intersecting null hyperplanes \emph{except} when ${\rm Re}\,\beta=0$ or ${\rm Im}\,\beta=0$ or ${\rm Re}\,\beta=c_0\,{\rm Im}\,\beta$.

\setcounter{equation}{0}
\section{Discussion}
The mathematical model of colliding plane impulsive gravitational waves described in \cite{BH} results in a space--time of constant curvature following the collision. A spin--off is a new form of the de Sitter or anti--de Sitter line elements in a coordinate system based on intersecting null hypersurfaces. The hypersurfaces have non--vanishing 
shear and expansion and thus they are quite complicated.  Nevertheless they merit the consideration in section 2 on account of their origin in a mathematical model with a 
clear physical interpretation. The null hyperplanes in space--times of constant curvature discussed in sections 3 to 5 are spin--offs of the Ozsv\'ath--Robinson--R\'ozga \cite{ORR} generalisation of the plane fronted gravitational waves with parallel rays to include a cosmological constant. An unusual feature of the line elements of de Sitter and anti--de Sitter space--times found in \cite{ORR} is that they involve three real--valued functions of a null coordinate. The geometrical construction described in detail in sections 3 and 4 (which has recently been exhibited in \cite{PAH}) provides a clear explanation of the origin of these functions and also explains the role the functions play in 
determining whether or not the null hyperplanes intersect.

%%%%%%%%%%%%%%%%%%%%%%%%

\end{document}